\documentclass[12pt, letterpaper]{article}

\pdfoutput=1

\usepackage{amsmath}
\usepackage{amsfonts}
\usepackage{amssymb}
\usepackage{graphicx}

\begin{document}

\begin{flushright} {\footnotesize YITP-20-21, IPMU20-0017}  \end{flushright}
\vspace{0.5cm}

\begin{center}

\def\thefootnote{\fnsymbol{footnote}}

{\Large\bf Bounds on the Horndeski Gauge-Gravity Coupling}
\\[0.7cm]
{Alireza~Allahyari$^{1}$ \footnote{alireza.al@ipm.ir}}, 
{Mohammad~Ali~Gorji$^{2}$ \footnote{gorji@yukawa.kyoto-u.ac.jp}}, 
{Shinji~Mukohyama$^{2,3}$ \footnote{shinji.mukohyama@yukawa.kyoto-u.ac.jp}}
\\[0.5cm]
{\small \textit{$^{1}$School of Astronomy, Institute for Research in Fundamental Sciences (IPM) \\P.~O.~Box 19395-5531, Tehran, Iran }}\\
 {\small \textit{$^2$Center for Gravitational Physics, Yukawa Institute for Theoretical Physics, Kyoto University, \\Kyoto 606-8502, Japan}}\\
 {\small \textit{$^3$Kavli Institute for the Physics and Mathematics of the Universe (WPI), \\The University of Tokyo Institutes for Advanced Study, The University of Tokyo, \\Kashiwa, Chiba 277-8583, Japan}}\\

\end{center}

\vspace{.8cm}

\hrule \vspace{0.3cm}


\begin{abstract}
The Horndeski gauge-gravity coupling is the leading non-minimal interaction between gravity and gauge bosons, and it preserves all the symmetries and the number of physical degrees of freedom of the standard model of particle physics and general relativity. In this paper we study the effects of the non-minimal interaction in astronomy and cosmology, and obtain upper bounds on the associated dimensionless coupling constant $\lambda$. From the modification of equations of motion of gauge bosons applied to compact astronomical objects, we find upper bounds $|\lambda| \lesssim 10^{88}$, $|\lambda| \lesssim 10^{75}$ and $|\lambda| \lesssim 10^{84}$ from a black hole shadow, neutron stars and white dwarfs, respectively. The bound $|\lambda| \lesssim 10^{75}$ that is deduced from neutron stars is the strongest and provides twenty orders of magnitude improvement of the previously known best bound on this parameter.  On the other hand, the effects of this term on modification of the gravitational Poisson equation lead to a weaker bound $|\lambda| \lesssim 10^{98}$. From the propagation of gravitational waves we also find $|\lambda| \lesssim 10^{119}$, which is even weaker. 
\end{abstract}
\vspace{0.5cm} \hrule
\def\thefootnote{\arabic{footnote}}
\setcounter{footnote}{0}


\newpage

\section{Introduction}
\label{section-1}

The Standard Model (SM) of particle physics describes interactions of elementary particles through the electromagnetic, weak and strong forces. The predictions of the model were accomplished by the experiments which made it one of the greatest triumphs in theoretical physics. The setup contains 19 parameters, and among them the mass of the Higgs boson is the last one which was recently measured~\cite{Aad:2012tfa,Chatrchyan:2012xdj}. On the other hand, General Relativity (GR) is very successful to explain the gravitational force in a wide range of scales from the cosmological and solar-system scales all the way down to the sub-millimeter scale with only 2 parameters, i.e. Newton's constant and the cosmological constant. Although the SM and GR work well in their domains of validity, we need to find a unified theory which describes all the four known forces in the nature. For instance, we need such a theory to explain the beginning of the universe in the standard Big Bang cosmology and also to understand gravity at very high curvature regime such as near the center of black holes. Indeed, appearance of the singularities in the context of GR is the signature for a new theory which is believed to be a quantum theory of gravity. However, we can consider interaction of gravity with the SM particles even in the absence of such a theory. In recent years, people widely studied the effects of gravitational non-minimal coupling of the Higgs field of the SM in the context of inflationary models \cite{Bezrukov:2007ep,Barvinsky:2008ia,Bezrukov:2008ej} (see also \cite{Rubio:2018ogq} for review). Similarly, one may consider interaction of gravity with other particles in the SM of particle physics. For gauge bosons, described by the Electroweak (EW) and Quantum Chromodynamics (QCD) sectors in the SM, the leading non-minimal interaction with gravity is given by the so-called Horndeski non-minimal term with the Lagrangian density \cite{Horndeski:1976gi}
\begin{equation}\label{Horndeski-term}
L^{\rm H}_i = \sum_a \Big( R F^a_{i,\mu\nu}F_{i,a}^{\mu\nu} - 4 R_{\mu\nu} F_{i,a}^{\mu\alpha} F^{a\nu}_{i,\,\,\alpha} + R_{\mu\nu\alpha\beta} F_i^{a\mu\nu} F_{i,a}^{\alpha\beta} \Big) \,,
\end{equation}
where Greek indices $\mu,\nu, ... = 0,1,2,3$ are spacetime indices, $a$ labels the fiber indices, and $i$ labels different gauge sectors of the SM so that $i=1$ and $i=2$ respectively correspond to the hyper $U(1)$ and the $SU(2)$ of the EW theory and that $i=3$ is associated to the QCD sector with a $SU(3)$ fiber. The ratios among the three terms of the above non-minimal interaction Lagrangian density for each gauge sector are uniquely determined so that the setup is free of the so-called Ostrogradsky higher derivative ghost. For the Higgs inflation where the Higgs boson, as inflaton, only couples non-minimally to the Ricci scalar, we do not need to care about the appearance of the Ostrogradsky ghost. However, Ostrogradsky ghost may arise if we consider either higher derivative terms of the Higgs field or terms that are higher order in curvature. For the gauge bosons, the situation is more restricted so that Ostrogradsky ghost arise even if we naively consider the term like $R\sum_{a}F^a_{i,\mu\nu}F_{i,a}^{\mu\nu}$. Therefore, we need to consider the combination (\ref{Horndeski-term}) which is free of Ostrogradsky ghost so that the number of physical degrees of freedom does not change by adding this term to the standard minimally-coupled system. 

Taking this interaction into account opens up new possibilities from the theoretical side and it has been the subject of focused studies ranging from the early universe to black hole solutions. For example, this interaction could lead to what is dubbed as \textit{HYM}-flation in the early universe \cite{Davydov:2015epx}. Previously the coupling of a photon field to the curvature has been considered in the context of magnetogenesis in \cite{Turner:1987bw} (see also \cite{Mukohyama:2016npi} for a more recent application of the non-minimal coupling). Further generalizations and the observational implications of this term are presented in Refs. \cite{Heisenberg:2018vsk,deFelice:2017paw,DeFelice:2016uil}. The other favoured area in such generalizations is the black hole solutions. The black hole solutions in the generalized theories have been the subject of the study in Ref. \cite{Heisenberg:2017hwb}. In the context of gravitational wave physics, this model exhibits interesting properties regarding propagation of gravitational waves. For example, the power spectrum of the primordial gravitational waves has been investigated in Ref. \cite{Feng:2016vyi}.

The Horndeski non-minimal coupling (\ref{Horndeski-term}) respects all the symmetries of the SM and GR, and thus there is no symmetry principle that forbids it. Also, the Horndeki term avoids Ostrogradsky ghost and preserves the number of physical degrees of freedom. Therefore the Horndeski non-minimal term should be present in the SM coupled to GR unless fine-tuned to zero by hand. It would couple gravity non-minimally to all gauge sectors of the SM, i.e. EW and QCD. We then should consider couplings between gravity and gauge bosons as $\lambda_{i} L^{\rm H}_i$ with $i=1,2,3$ ($i=1$ for hyper $U(1)$, $i=2$ for $SU(2)$ and $i=3$ for $SU(3)$, respectively). Therefore, although the SM has 19 parameters and GR has 2 parameters, the combination of the two theories would have more parameters such as $\lambda_i$, which show up due to the interaction of gravity and SM particles. The 19 parameters of SM and 2 parameters of GR are measured precisely by many experiments. Although it is not easy to measure $\lambda_i$ within currently accessible energy and curvature scales, we can at least put some upper bounds on these parameters since we know that SM and GR work well in their domains of applicabilities. More precisely, we should take into account the effects of the Horndeski non-minimal term in both the SM and GR. For instance, gravity is so weak and we can safely neglect the Einstein-Hilbert term and also the Horndeski non-minimal coupling term in comparison with the free Yang-Mills term in the standard calculations of the QED and QCD. Nonetheless, we note that we can still find some bounds on $\lambda_i$ simply from the fact that the EW and QCD predictions without the non-minimal coupling match very well with experiments. In the same manner, we can consider its effects when gravity couples to any gauge field like the radiations in our universe and we find that it is small while again we get some bounds on $\lambda_i$. Depending on the setup that one chooses to explore for the effects of this non-minimal coupling, weaker or stronger bounds on $\lambda_i$ may arise. The aim of this paper is to find these types of setup and, then, find some upper bounds on the non-minimal coupling parameters $\lambda_i$.

The rest of the paper is organized as follows. In Section \ref{section-2}, we introduce the theoretical setup and we find general statements which we need to constrain the Horndeski non-minimal term. In Section \ref{section-3}, looking for possible modification of the QED and QCD sectors of SM, we study compact astronomical objects such as black holes, neutron stars and white dwarfs. We find the tightest bound in the case of neutron stars. In Sections \ref{section-4} and \ref{section-5}, we look at the modifications of the gravitational Poisson equation and the propagation of gravitational waves induced by the Horndeski non-minimal term and we find bounds which are weaker than what we find from the compact astronomical objects. Section \ref{section-6} is devoted to the summary of the results.

\section{The setup}
\label{section-2}

The free action for gravity is given by the Einstein-Hilbert term and the free action for  gauge bosons is given by the Yang-Mills action. We also consider interaction between gravity and gauge bosons and, therefore, it is natural to consider the following Einstein-Yang-Mills system
\begin{eqnarray}\label{action}
S = \frac{M_P^2}{2} \int d^4x \sqrt{-g} (R-2\Lambda) + \sum_i S_{i}\,, \quad 
S_i = \int d^4x \sqrt{-g} \bigg[ 
  - \frac{1}{4} \sum_{a} F^a_{i,\mu\nu}F_{i,a}^{\mu\nu} 
  - \frac{\lambda_{i}}{4 M_P^2} L^{\rm H}_{i} \bigg] \,,
\end{eqnarray}
where $M_P = (8{\pi}G)^{-\frac{1}{2}}$ is the reduced Planck mass and $\Lambda$ is the cosmological constant\footnote{We work in the units $\hbar=1=c$, where $\hbar$ is the reduced Planck constant and $c$ is the speed of light in vacuum.}. The action is gauge-invariant and also provides a set of second order equations of motion which ensures the absence of Ostrogradsky ghost \cite{Jimenez:2013qsa,BeltranJimenez:2017cbn}. The non-minimal parameters $\lambda_i = {\mathcal O}(1)$ label the Horndeski term defined in (\ref{Horndeski-term}) which is the leading non-minimal interaction for Einstein-Yang-Mills system. As we mentioned above, the gauge bosons labeled by $i$ are those of the EW and QCD sectors of the SM. There are, of course, some other interactions which we have omitted. In each gauge boson sector, there are some interactions which are present in the SM and the associated coupling constants are also precisely measured. In the gravity sector, there are Newton's constant and the cosmological constant which are precisely measured. There would be also some higher derivative self-interactions terms like $R^2$ which are constrained by solar, astronomical, and cosmological observations. In this regard, the Horndeski term (\ref{Horndeski-term}) is the leading non-minimal interaction for the Einstein-Yang-Mills system. Both the equations of motion of gauge bosons and gravitons get modified by the Horndeski non-minimal interaction. Since we demand that the Horndeski non-minimal interaction term be sufficiently smaller than the Einstein-Hilbert and Yang-Mills terms, we can treat the effects of the Horndeski term iteratively. This makes the analysis very simple as follows. In the absence of the Horndeski interaction, we know how to analyze solutions for both the gauge and gravity sectors determined by the Yang-Mills and Einstein's-Hilbert terms respectively. Having these solutions in hand, we can substitute them into the Horndeski non-minimal interaction and estimate its order of magnitude for the system under consideration. More precisely, we need to demand that this value be smaller than the both Yang-Mills and Einstein's-Hilbert terms. Thus, we find some upper bounds on the non-minimal parameters $\lambda_i$. This is our strategy to constrain $\lambda_i$ in this paper. For finding these types of estimations, we really do not need to consider the explicit form of the Horndeski term shown in (\ref{Horndeski-term}). We define the order of magnitude ${\cal R}$ of curvature as a general quantity that could be constructed linearly from the Ricci scalar, the Ricci tensor or Riemann tensor in tetrad basis. We will explicitly find this quantity in two cases of compact astronomical objects and an expanding cosmological background in this paper which makes clear the notion of this quantity. Thus, the Horndeski non-minimal interaction term in the action (\ref{action}) can be schematically written as 
\begin{equation}\label{Horndeski-term-G}
\lambda  L^{\rm H} \sim \frac{\lambda}{M_P^2} {\cal R} {\cal F}^2 \,,
\end{equation}
where depending on the system under consideration, $\lambda$ could be one of the non-minimal coupling parameters $\lambda_i$ ($i=1,2,3$) defined in (\ref{action}) for the hyper $U(1)$, $SU(2)$ and $SU(3)$ sectors and ${\cal F}$ is the order of magnitude of the corresponding field strength tensor in tetrad basis. In the case of electrodynamics it is given by ${\cal F} \sim \max (E, B)$, where $E$ ($\geq 0$) and $B$ ($\geq 0$) are the order of magnitudes of the electric and magnetic fields.

As we mentioned above, the Horndeski interaction term (\ref{Horndeski-term-G}) should be smaller than both the Einstein-Hilbert and Yang-Mills terms in the action (\ref{action}). The latter implies $\frac{\lambda}{M_P^2} {\cal R} \lesssim 1$. Considering ${\cal R}$ as an average Gaussian curvature, we can define the curvature length scale (or radius) $\ell_{\cal R}$ as ${\cal R} \equiv {\cal O}(1) {\ell_{\cal R}}^{-2}$. Then, we find the following condition
\begin{equation}\label{lambda-condition-R}
|\lambda| \lesssim \Big( \frac{\ell_{\cal R}}{\ell_P} \Big)^2 \,,
\end{equation}
where $\ell_P=M_P^{-1}$ is the reduced Planck length. On the other hand, demanding that the Horndeski interaction term (\ref{Horndeski-term-G}) be smaller than the Yang-Mills term, we also find
\begin{equation}\label{lambda-condition-F}
|\lambda| \lesssim \bigg( \frac{M_P^2}{{\cal F}} \bigg)^2 \,.
\end{equation}	

The condition (\ref{lambda-condition-R}) guaranties that corrections to the equation of motion of gauge bosons are small and condition (\ref{lambda-condition-F}) guaranties that corrections to the equation of motion of gravitons are small. These requirements are rather conservative but still lead to meaningful upper bounds on $|\lambda|$. The first condition (\ref{lambda-condition-R}) has been implemented in Ref.~\cite{Barrow:2012ay} and they found $|\lambda|\ll 10^{90}$ for earth-based experiments. Finding these types of bounds crucially depends on two different sides of the experiment under consideration, the accuracy of measurement in the experiment and the strength of gravity/gauge force. The more accurate measurements and stronger gravity/gauge force regimes will lead to tighter bounds on $\lambda$. Measurements on the earth are usually more accurate than other measurements such as those at the solar system scale and also in cosmology. Nonetheless, in principle, we can still find tighter bounds by means of less accurate measurements but with strong gravity/gauge force. This is what we look for in this paper and, indeed, we find a tighter bound from consideration of compact astronomical objects, which is the subject of the next section.

\section{Modifications of gauge forces in SM}
\label{section-3}

From (\ref{lambda-condition-R}), we see that in order to obtain tighter bounds on the Horndeski non-minimal coupling parameter $\lambda$, we need to look for objects with small curvature radius $\ell_{\cal R}$. In the case of astronomical objects such as stars and black holes, as we will show, the curvature scale is determined by the radius and mass of the object. In this section, we first find an expression for the curvature radius of the astronomical objects. Using the result, we look for modification to the equation of motion of gauge bosons  coming from the Horndeski term in black hole shadows, neutron stars, and white dwarfs as environments which include gauge bosons. We then find some upper bounds on $\lambda$ accordingly.

The metric around a compact astronomical object like black holes, neutron stars, white dwarfs, and usual stars, can be approximately described by the spherically symmetric Schwarzschild background
\begin{equation}\label{schwarzschild-metric}
ds^2_{\rm S} = - \Big( 1-\frac{r_{\rm S}}{r} \Big) dt^2 
+ \Big(1-\frac{ r_{\rm S}}{r}\Big)^{-1} dr^2 + r^2 d\Omega^2 \,, 
\end{equation}
where $r_{\rm S} = \frac{M}{4\pi M_P^2}$ is the so-called Schwarzschild radius. In the case of black holes, $M$ is the mass of the black hole and $r_{\rm S}$ is the horizon radius while for a star, $M$ is the mass of the star but the radius located at $r=r_\ast$ is larger than the Schwarzschild radius $r_\ast>r_{\rm S}$. Realistic astronomical objects are neither spherically symmetric nor in vacuum, but the Schwarzschild metric is still useful for the order of magnitude estimate of the curvature ${\cal R}$. For the sake of simplicity we also neglect the effects of rotation. A better exterior metric for astronomical objects is given in Ref. \cite{HaTh}.

The Ricci scalar and Ricci tensor for the Schwarzschild metric (\ref{schwarzschild-metric}) vanish but the Riemann tensor does not vanish. We, therefore, can use the Riemann tensor to find the order of magnitude of curvature ${\cal R}$. The non-zero tetrad components of the Riemann tensor are of order $r_{\rm S}/r^3$, and therefore we would have ${\cal R} \sim r_{\rm S}/r^3$ which using ${\cal R} = {\cal O}(1){\ell_{\cal R}}^{-2}$ gives $\ell_{\cal R} \sim r^{3/2}/\sqrt{r_{\rm S}}$. This can be also concluded from the Kretschmann scalar as ${\cal R} \sim \sqrt{R_{\mu\nu\alpha\beta}R^{\mu\nu\alpha\beta}}$. Writing explicitly in terms of the radial coordinate $r$ and mass, we find the curvature radius for a astronomical object with spherical symmetry as follows
\begin{equation}\label{curvature-radius-Schw}
\ell_{\cal R} \sim \frac{M_P}{\sqrt{M}} r^\frac{3}{2} \,.
\end{equation}

From (\ref{lambda-condition-R}), we see that the smaller curvature length scale we achieve, the tighter bound on $\lambda$ we find. In this regard, we need to estimate this relation at some high curvature regime to find tight bounds on $\lambda$.

\subsection{Black hole shadows}

From (\ref{lambda-condition-R}) we see that we need objects with small curvature radii, i.e. with strong gravity, to find tight bounds on $\lambda$. Therefore black holes are of our interest. There are wealth of evidences hinting the existence of supermassive black holes with masses as large as $10^{10} M_{\odot}$ where $M_{\odot}$ is the solar mass. The convention is that supermassive black holes reside in the center of sufficiently massive galaxies like Milky way \cite{LyndenBell:1969yx, Kormendy:1995er}. Apart from the gravitational part of the system which is a black hole here, we need physical processes in which gauge bosons play major roles. Fortunately, the photons in the context of black hole shadows can be considered as the desired gauge boson in our setup. The shadow of a black hole was first studied by Synge \cite{Synge:1966okc}. This result was later extended to a Kerr black hole \cite{Bardeen}. The shadow is the boundary of photons path that when traced back from the observer, reach the unstable photon orbit of the black hole \cite{Dokuchaev:2019jqq}. For the first time, the observation of a black hole shadow has become a reality through the Event Horizon Telescope (EHT) collaboration \cite{Akiyama:2019cqa,Akiyama:2019eap,Akiyama:2019fyp,Akiyama:2019bqs,Akiyama:2019sww,Akiyama:2019brx}. We can therefore study the effects of the presence of the Horndeski non-minimal term on the shadows of black holes. 

The size of a shadow is roughly determined by the impact parameter $b=L/E$, where $L$ and $E$ denote the energy and the angular momentum respectively. For a Schwarzschild black hole with the metric (\ref{schwarzschild-metric}), the size is given by\footnote{In fact, photons follow geodesics of an effective metric in the geometric optics approximation \cite{Allahyari:2019jqz}.}
\begin{equation}\label{b}
b \sim 3\sqrt{3}\frac{M}{M^2_P} \,.
\end{equation}

The presence of the Horndeski non-minimal coupling changes the equation of motion of the photons. To estimate the order of magnitude of the effects of the Horndeski term, we do not need to go for the explicit calculations. The equations of motion for the photons are determined by the two terms in $S_i$ in (\ref{action}). Therefore, the order of magnitude of the corrections would be obtained by comparing these terms with each other. Following our discussions in section \ref{section-2}, the ratio of the Horndeski term to the standard kinetic term is of the order of $\lambda{\cal R}/M_P^2 = \lambda (\ell_P/\ell_{\cal R})^2$. Taking this fact into account and by dimensional analysis we find that the correction to the impact parameter $b$ of a black hole shadow will be 
\begin{equation}\label{impact-parameter}
\frac{\delta b}{b} \sim \lambda \Big( \frac{\ell_P}{\ell_{\cal R}} \Big)^2 \,,
\end{equation}
where $\delta{b}$ is the variation of the impact parameter due to the existence of the Horndeski non-minimal term. To respect the observed impact parameter of the black hole shadow, this correction should be less than unity which gives the condition (\ref{lambda-condition-R}) that we already obtained. More precisely, these corrections should be smaller than the accuracy of the measurement. However, the measurement of the shadow of $M87^\ast$ black hole by EHT is not very accurate and the accuracy is actually of order unity, i.e. $|\delta{b}/b| \lesssim {\cal O}(1)$. By combining this with (\ref{impact-parameter}) we would obtain an upper bound on $|\lambda|$. Therefore, (\ref{lambda-condition-R}) is a good estimate for the black hole shadow. We then just need to find the curvature radius $\ell_{\cal R}$ for this case. The shadow is located approximately around the photon sphere $r_{\rm b} \approx \frac{3\sqrt{3}}{8\pi} \frac{M_{\rm bh}}{M_P^2}$ where $M_{\rm bh}$ is the black hole mass. Substituting $r \sim r_{\rm b}$ to (\ref{curvature-radius-Schw}), we find
\begin{equation}\label{curvature-radius-Schw-shadow}
\ell_{\cal R} \sim \frac{M_{\rm bh}}{M_P^2} \,,
\end{equation}
which after substituting to (\ref{lambda-condition-R}) gives
\begin{equation}\label{lambda-condition-BHs}
|\lambda| \lesssim \Big(\frac{M_{\rm bh}}{M_P}\Big)^2 \,.
\end{equation}

The above relation can be used for the shadow of any black hole to find upper bounds on the Horndeski non-minimal parameter $\lambda$ in the case of the QED sector since we deal with photons here. The photon sphere of a black hole, which defines the curvature radius, is completely determined by the mass of the black hole and, therefore, the mass of the black hole is the only parameter of the object which we deal with. From (\ref{lambda-condition-BHs}), we see that we can in principle consider black holes with smaller masses which are more compact to find tighter bounds on $\lambda$. We, however, have not observed shadows of black holes with small masses. Let us therefore apply the setup to the observed supermassive black hole $M87^{*}$ with large mass
\begin{eqnarray}\label{shadow-mass}
M_{\rm bh} = M^{87^{*}}_{\rm bh} \simeq 10^{6} M_{\odot} \,.
\end{eqnarray}
From (\ref{lambda-condition-BHs}), we then find the following bound
\begin{equation}\label{lambda-shadow}
|\lambda| \lesssim 10^{88} \,,
\end{equation} 
where we have substituted $M_{\odot} \simeq 10^{66} \mbox{eV}$ for the solar mass. We may try to take into account the accuracy of the experiment but we find that $\frac{\delta b}{b} \leq 1.5$ for $M87^{*}$ \cite{Akiyama:2019eap,Akiyama:2019brx} and thus the above bound essentially does not change. The above bound can be improved with the future observations with higher accuracy. 

The bound (\ref{lambda-shadow}) is not better than the bound $|\lambda| \ll 10^{90}$ that was obtained in Ref. \cite{Barrow:2012ay} based on the experimental data on the earth. One might have naively expected that a supermassive black hole with huge mass of (\ref{shadow-mass}) might lead to a better bound. However, the point is that the compactness of an astronomical object is the criterion to get tighter bounds which can be understood from the expression of the curvature radius (\ref{curvature-radius-Schw}). One  point  that is missing in our treatment of the  shadow of $M87^{*}$ is the effect of rotation. We assumed a spherically symmetric, static black hole. The effect of rotation when  taken into account is to introduce asymmetry in the shadow \cite{Bambi:2019tjh}. However, it is expected that this does not significantly change our order of magnitude estimation. The effects of rotation in the presence of the non-minimal interaction term could nonetheless be treated in a future work.

\subsection{Neutron stars} 

We now consider neutron stars which are the remnants of stars after supernova explosions. They are among the most compact objects in the universe and we expect to obtain a tight bound since from (\ref{lambda-condition-R}) we know that compactness is the key criterion for tighter bounds.

Properties of neutron stars depend on the nucleon-nucleon interaction which is described by the QCD sector of the SM. Therefore, the gauge bosons in this case are gluons in QCD. We can then use the formula (\ref{lambda-condition-R}) for QCD and neutron stars. We first note that the QCD gauge bosons non-minimally couple to the gravity through the Horndeski term which changes the effective gauge coupling constant. Quantities that are usually constant in nuclear physics such as the neutron mass will change as a result. Properties of neutron stars such as masses and radii depend on the nuclear physics \cite{Shapiro:1983du,Ishak:2018his} and therefore will change accordingly. If the effective gauge coupling constant received corrections of order unity, then the modified QCD prediction of neutron star properties would be inconsistent with observations. Therefore, to keep essentially the same physics for neutron stars as in the standard QCD, we demand that this kind of change is not as large as order unity and then we can use the formula (\ref{lambda-condition-R}) to constrain the non-minimal coupling parameter for the QCD sector.

We first need to find the curvature radius. Approximating the exterior geometry around a neutron star by the Schwarzschild metric (\ref{schwarzschild-metric}), the curvature radius is given by (\ref{curvature-radius-Schw}). We need the mass and radius of a neutron star and the typical values are given by
\begin{eqnarray}\label{mass-radius-n}
M_{\rm n} \simeq 1.5 M_{\odot} \,, \hspace{1cm} r_{\rm n} \simeq 1.4 \times 10^{-5} r_{\odot} \,,
\end{eqnarray}
where $r_{\odot} \simeq 3.5 \times 10^{15} \, \mbox{eV}^{-1}$ is the solar radius.

Substituting $M = M_{\rm n}$ and $r = r_{\rm n}$ to (\ref{curvature-radius-Schw}) and then using the result in (\ref{lambda-condition-R}), we find
\begin{equation}\label{lambda-neutronstar}
|\lambda| \lesssim 10^{75} \,.
\end{equation} 
As far as we know, this is the tightest bound on the Horndeski non-minimal coupling parameter that has ever been found. The reason can be understood if we note that the curvature radius (\ref{curvature-radius-Schw}) for a typical neutron star with mass and radius (\ref{mass-radius-n}) is really small. Then, from (\ref{lambda-condition-R}) we obtained the tight bound (\ref{lambda-neutronstar}).

\subsection{White dwarfs}

The object we consider in this subsection is a white dwarf which is composed of degenerate electron gas. The balance of gravitational pull and degeneracy pressure determines the mass of a white dwarf \cite{Shapiro:1983du}. If the corrections to the proton mass due to the Horndeski non-minimal coupling is of order unity then the gravitational pull due to the total mass of protons significantly changes for a given number of electrons (which is equal to the number of protons) and as a result properties of white dwarfs such as masses and radii would become inconsistent with observations. Therefore again we can use (\ref{lambda-condition-R}) to obtain an upper bound on $|\lambda|$ for the QCD sector of the SM.  Although the equation of state of degenerate electron gas slightly depends on the electromagnetic interaction and weak interaction, the main dependence comes from the degeneracy pressure which is determined by the Pauli principle and thus is basically independent of gauge forces. Therefore, from white dwarfs the constraint on $\lambda$ in the EW sector of the SM is expected to be weaker than in the QCD sector and we shall not consider the former in this paper.

To find the bound, we need typical mass and radius of a white dwarf. There are astronomical observations which determine the mass and radius of white dwarfs. Here our aim is to use these observations to constrain the Horndeski non-minimal coupling parameter. These types of observations are also used to constrain other fundamental constants \cite{Landau:2020vkr,Magano:2017mqk}. The typical values for the mass and radius of a white dwarf are 
\begin{eqnarray}\label{mass-radius-w}
M_{\rm w} \simeq 1.2 M_{\odot} \,, \hspace{1cm} r_{\rm w} \simeq 10^{-2} r_{\odot} \,.
\end{eqnarray}

Comparing the above values with those for a neutron star in (\ref{mass-radius-n}), we see that the mass of a white dwarf is the same order as a neutron star while the radius is larger by a factor of $10^3$. Therefore, bounds coming from white dwarfs will be weaker than the bounds that we obtained from neutron stars. From (\ref{curvature-radius-Schw}) and (\ref{lambda-condition-R}) we then find the following bound
\begin{equation}\label{lambda-whiteduarf}
|\lambda| \lesssim 10^{84} \,.
\end{equation}

Although the above bound is not as good as the bound (\ref{lambda-neutronstar}), it is still much better than bound (\ref{lambda-shadow}) which we obtained from the shadow of $M87^\ast$ suppermassive black hole and the bound $\lambda\ll 10^{90}$ that was obtained in Ref. \cite{Barrow:2012ay} from the earth-based experiments. This shows that after neutron stars, white dwarfs are the best astronomical candidates to constrain the Horndeski non-minimal parameter $\lambda$.

\subsection{Summary of bounds from modification of gauge forces}

\begin{figure}[htbp!]
	\centering
	\includegraphics[width=0.5 \columnwidth]{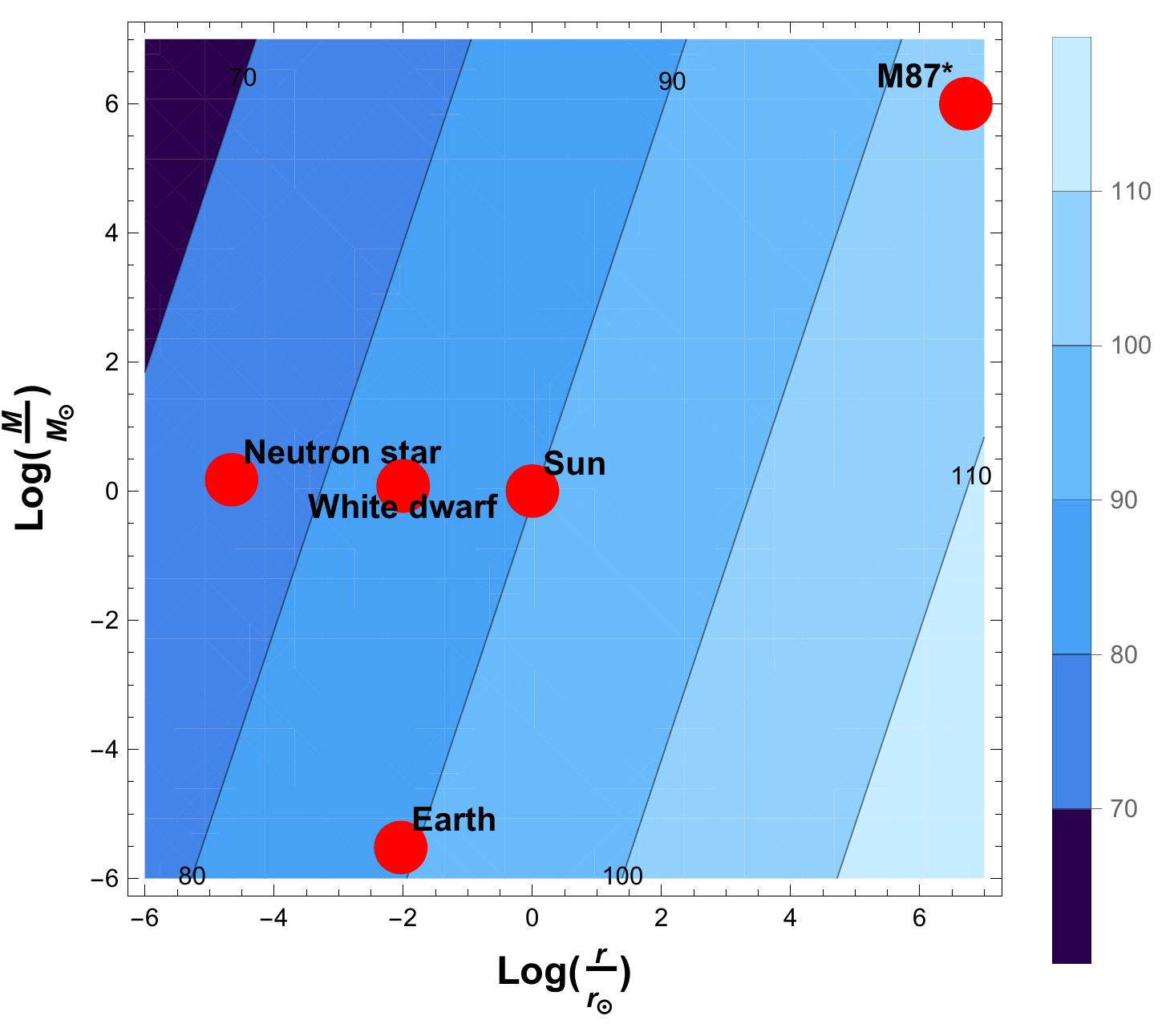}
	\caption{Bounds on $|\lambda|$ from different observations as a function of $\log\left(M/M_{\odot} \right) $ and $\log\left( r/r_{\odot}\right) $. The value on each contour represents the corresponding upper bound on $\log_{10}|\lambda|$.}
	\label{fig:1}
\end{figure}

Let us summarize the results for different observations in Fig.~\ref{fig:1}. In this figure, we have illustrated estimations for the upper bounds on $\log_{10}|\lambda|$ as a function of $\log\left(M/M_{\odot} \right) $ and $\log\left( r/r_{\odot}\right) $ in terms of color contours. The value on each contour represents the upper bound on $\log_{10}|\lambda|$.

From the condition (\ref{lambda-condition-R}), we see that in order to find tighter bounds, we need to consider objects with smaller curvature length scales. For the astronomical objects, the curvature length scale is given by (\ref{curvature-radius-Schw}) which shows that we need smaller radius and larger mass to achieve smaller curvature length scale. The supermassive $M87^\ast$ black hole has the largest mass while the earth has the smallest one. On the other hand, neutron stars have the smallest radius while the $M87^\ast$ black hole has the largest radius. Naively, one might expect to find a very tight bound from a black hole since gravity is the dominant force controlling the system. However, as we can see in Fig.~\ref{fig:1}, supermassive $M87^\ast$ black hole gives the weakest bound among those we considered and it is even weaker than one from the earth. The reason is that although the $M87^\ast$ black hole has the largest mass, it has the largest radius while we need smaller radius to achieve small curvature length scale. For the sun and the earth, we obtain essentially the same bounds since the curvature length scale given by (\ref{curvature-radius-Schw}) becomes almost the same for them. From Fig.~\ref{fig:1} we also see that neutron stars, white dwarfs and the sun appear almost on the same horizontal line as the masses of these objects are almost the same (see (\ref{mass-radius-n}) and (\ref{mass-radius-w})). However, the radii of neutron stars are smaller than white dwarfs and the sun. Therefore, we obtain the tightest bound from neutron stars. Similarly, white dwarfs give a bound stronger than one from the sun since their radii are smaller.

\section{Modification of gravitational Poisson equation}
\label{section-4}

Up to here, we always looked at the equations of motion of the gauge bosons and, therefore, we deal with condition (\ref{lambda-condition-R}). As we mentioned in Section \ref{section-2}, the equation of motion of the metric also gets modified in this scenario and we found another condition (\ref{lambda-condition-F}). 

One way to look for such a modification is to note that the effects of Horndeski non-minimal term in the action (\ref{action}) can be understood as a shift of the effective Newton's constant through the corresponding modified Poisson equation. Using the results of Section \ref{section-2}, the order of magnitude of this correction can be estimated by comparing the Einstein-Hilbert term with the Horndeski non-minimal term as follows
\begin{equation}\label{Newton-constant}
\frac{\delta{G}}{G} \sim \lambda \bigg( \frac{{\cal F}} {M_P^2}\bigg)^2 \,.
\end{equation}

From the above relation, we see that in order to obtain an upper bound, we need to consider a system with nonzero ${\cal F} \sim \max(E,B)$. The electric and magnetic fields are non-zero even on the earth and we can then find some bounds. Restricting to the case of magnetic field (and thus neglecting electric field) and then demanding that corrections coming from Horndeski non-minimal term be smaller than the observational upper bound on the variation of Newton's constant, we find
\begin{equation}\label{lambda-Newton-constant}
 |\lambda| \lesssim \left|\frac{\delta{G}}{G}\right|_{\rm bound} \, \bigg( \frac{M_P^2}{B}\bigg)^2 \,.
\end{equation}
where $|\delta G/G|_{\rm bound}$ is the observational upper bound on $|\delta G/G|$. 

From the above relation, we see that tighter bounds come from stronger magnetic fields. Magnetic fields are present in standard stars, white dwarfs, neutron stars, and magnetars. The magnetic field on the sun and the earth is $B\sim 1 \mbox{G}$ and $B\sim 0.5 \mbox{G}$ respectively. The magnetic field on a white dwarf could be as strong as $B\sim 10^{6} \mbox{G}$. Magnetars could have magnetic fields of order $B\sim 10^{15} \mbox{G}$. For white dwarfs and neutron stars, it is believed that there is a balance between the gravitational force and the interior  degeneracy pressure. If $\lambda$ is too large this could violate the known physics of these objects by changing the gravitational force. This modifies the white dwarf mass-radius relation and the neutron star mass limit as these relation and limit depend on the gravitational Poisson equation \cite{Jain:2015edg}.

To get stronger bound in this case, we consider neutron stars with typical magnetic field $B\approx10^9$ $\mbox{G}$. Noting that $1\, \mbox{G} = 1.95 \times 10^{-20}$ $\mbox{GeV}^2$, we find
\begin{align}
|\lambda| \lesssim 10^{98} \,, 
\end{align}
which is weaker than all of the bounds (\ref{lambda-shadow}), (\ref{lambda-neutronstar}), and (\ref{lambda-whiteduarf}) that we deduced from modification of gauge forces for the compact astronomical objects.

\section{LIGO bound}
\label{section-5}

The Horndeski non-minimal interaction in the action (\ref{action}) slightly changes the propagation of gravitational waves (GWs) and electromagnetic waves as well. We have also seen in the previous section that the gravitational Poisson equation is also modified. The possibility of variation of fundamental constants and their implications are vast topics \cite{Uzan:2010pm}. 

In order to study propagation of GWs, we consider cosmological metric
\begin{equation}\label{FLRW-metric}
ds_{\rm C}^2 = - dt^2 + a(t)^2 (\delta_{ij} + h_{ij}(t,x) ) dx^i dx^j \,,
\end{equation}
where $t$ is the cosmic time, $a(t)$ is the scale factor, and $h_{ij}(t,x)$ are transverse traceless $\partial_i h_{ij} = 0 = h_{ii}$ tensor perturbations which encode GW polarizations. Taking variation of the action (\ref{action}) with respect to the metric, up to the linear order, we find the following form of the equation of motion for the GWs
\begin{equation}\label{GW-EoM}
\ddot{h}_{ij} + 3 H \dot{h}_{ij} - c_{\rm T}^2 \nabla^2 h_{ij} + m_{\rm g}^2 h_{ij} = 0 \,,
\end{equation}
with $c_{\rm T} = 1$ and $m_{\rm g} = 0$, where $H \equiv \dot{a}/a$ is the Hubble parameter. In general the parameter $c_{\rm T}$ defines the speed of GWs and parameter $m_{\rm g}$ defines the mass of gravitons. Introducing a non-zero background magnetic field, the equation of motion is modified. In general homogeneity and isotropy are broken already at the level of the background and thus the modified equation of motion for GWs becomes rather complicated and no longer of the form (\ref{GW-EoM}). Nonetheless, for the order of magnitude estimations it is sufficient to consider the modification in the form of the effective speed and the effective mass of GWs. Rigorously speaking, even these quantities become direction- and position-dependent but we do not need to consider their explicit forms for our purpose.

Let us first look at the corrections to the speed of GWs. To get some bounds on $\lambda$ from the GWs observations, we should compare the speed of GWs with the speed of electromagnetic waves. In our setup, both the speed of GWs and speed of electromagnetic waves get modified. We therefore need to find corrections to  both. Looking at the action (\ref{action}), we find that these corrections would be of order
\begin{equation}\label{GWs-speed}
\frac{\delta {c}_{\rm T}}{{c}_{\rm T}} \sim \lambda \bigg( \frac{ {\cal F}}{M^2_P} \bigg)^2 \,, \hspace{1.5cm}
\frac{\delta c_{\gamma}}{c_{\gamma}} \sim \lambda\frac{\cal R}{M^2_P} \,,
\end{equation}
where $c_{\gamma}$ is the speed of electromagnetic waves while $\delta{c_{\rm T}}$ and $\delta{c}_\gamma$ show small deviations induced by the Horndeski term. Now, we need to determine ${\cal F}$ and ${\cal R}$ in this case. As we mentioned in the last section, the magnetic field is present in the cosmological background and around astronomical objects and therefore we consider ${\cal F} \sim B$. In the case of curvature scale ${\cal R}$, we note that in the cosmological background, the Ricci scalar does not vanish and we can use it to define ${\cal R}$. Using the Ricci tensor and Riemann tensor components give the same results. In the cosmological background (\ref{FLRW-metric}), we have $R \sim H^2$ and therefore we consider ${\cal R} \sim H^2$. Although this curvature scale is time-dependent, the cosmological time scale is sufficiently long and thus we can still approximately define curvature length scale as
\begin{equation}\label{curvature-length-cosmology}
\ell_{\cal R} \sim H^{-1} \,,
\end{equation}
which is nothing but the Hubble horizon radius for the cosmological background (\ref{FLRW-metric}). The corrections (\ref{GWs-speed}) can then be rewritten as
\begin{equation}\label{GWs-speed-f}
\frac{\delta {c}_{\rm T}}{{c}_{\rm T}} \sim \lambda \Big( \frac{B}{M^2_P} \Big)^2 \,, \hspace{1.5cm}
\frac{\delta c_{\gamma}}{c_{\gamma}} \sim \lambda \Big(\frac{H}{M_P}\Big)^2 \,.
\end{equation}

We therefore would have

\begin{equation}\label{GW-corrections0}
\frac{| {c}_{\rm T} - c_{\gamma} |}{c_\gamma}=\frac{|\delta {c}_{\rm T}-\delta c_{\gamma}|}{c_{\gamma}} \sim \lambda \times
\mbox{min} \bigg[ \Big( \frac{B}{M^2_P} \Big)^2 , \Big(\frac{H}{M_P}\Big)^2 \bigg] \,.
\end{equation}
To find the minimum, we compute the ratio of the two corrections as
\begin{equation}\label{speed-EM-GW}
\frac{\delta{c}_{\rm T}/{c}_{\rm T}}{\delta c_{\gamma}/c_{\gamma}} 
\sim \Big(\frac{B}{HM_P}\Big)^2 \,.
\end{equation}

The present value of the Hubble parameter is $H_0 \simeq 10^{-33}$ $\mbox{eV}$ which gives a very large radius after using (\ref{curvature-length-cosmology}). The range of background magnetic field at cosmological scales is $10^{-15} \mbox{G} <B<10^{-9} \mbox{G}$. Using these values in the above formula we find that $\frac{\delta{c}_{\rm T}/{c}_{\rm T}}{\delta c_{\gamma}/c_{\gamma}} \sim 10^{7} - 10^{13}$ which means that we can neglect corrections to the speed of electromagnetic waves in comparison with the corrections to the speed of GWs. Using this result in (\ref{GW-corrections0}), we have
\begin{equation}\label{GW-corrections}
\frac{| {c}_{\rm T} - c_{\gamma} |}{c_\gamma} \sim \lambda \Big(\frac{B}{M^2_P} \Big)^2 \,.
\end{equation}

From the above relation, we find the following bound on the Horndeski non-minimal coupling parameter
\begin{equation}\label{lambda-GW0}
|\lambda| \lesssim \Big{|} 1 - \frac{{c}_{\rm T}}{c_\gamma} \Big{|}_{\rm bound} \Big( \frac{M^2_P}{B} \Big)^2 \,,
\end{equation}
where $|1-c_{\rm T}/c_{\gamma}|_{\rm bound}$ is the observational upper bound on $|1-c_{\rm T}/c_{\gamma}|$. Comparing the above relation with (\ref{lambda-condition-F}), we see that the above bound is much better since from GWs observations we have $\Big{|} 1 - {c}_{\rm T}/c_\gamma \Big{|}\lesssim 10^{-15}$. Taking this factor to be of order unity is the minimal assumption to get bound which we assumed in Section \ref{section-2} to find some general criterions like (\ref{lambda-condition-R}) and (\ref{lambda-condition-F}). However, as we can see from (\ref{lambda-GW0}), the accuracy of the measurement in the experiment can significantly improve the bound. In the case of GWs, substituting the upper bound for the background magnetic field $B \sim 10^{-9} G$ gives
\begin{equation}\label{lambda-GW}
|\lambda| \lesssim 10^{119} \,,
\end{equation}
which is weaker than the bounds that we obtained from modification of gauge forces for astronomical objects (\ref{lambda-shadow}), (\ref{lambda-neutronstar}), (\ref{lambda-whiteduarf}), and also than the bound (\ref{lambda-Newton-constant}) that we obtained from modification of gravitational Poisson equation. 

Now, we estimate the effective mass term that appears in the equation of motion of GWs (\ref{GW-EoM}). The leading effective mass term provided by the Horndeski non-minimal term is given by $m^2_{\rm g} \sim (\lambda B/M_P)^2$ which is small compared to the effective mass term provided by the free Yang-Mills term that is of order $\sim B^2$. In this respect, we do not obtain a significant bound on $\lambda$, since the effective mass term provided by the Horndeski non-minimal coupling term is suppressed in comparison with the effective mass term induced by the free Yang-Mills term.

\section{Summary}
\label{section-6}

The Horndeski gauge-gravity coupling is the leading non-minimal interaction of gravity with gauge bosons such as those in the EW and QCD sectors of the standard model (SM) of elementary particles. In weak gravity environments, this term is subdominant and one can usually neglect its effects. However, it would play significant roles in the high curvature regime for any gravitational system that include any types of gauge bosons unless the corresponding coupling constant is sufficiently small. It is therefore important and interesting to explore cosmological and phenomenological implications of the non-minimal coupling. Then, the first step in this direction is to estimate the order of magnitude of this term in physical systems. In the present paper we have studied the effects of the Horndeski non-minimal term in astronomy and cosmology and have found some upper bounds on the corresponding dimensionless non-minimal parameter $\lambda$. We first considered compact astronomical objects such as black holes, neutron stars and white dwarfs. In the case of black holes, the shadow is constructed from photons and black hole is the corresponding gravitational system. Therefore, we considered the effects of Horndeski term on the propagation of the photons. For the supermassive black hole $M87^\ast$, the EHT recently detected the shadow and we used this observation to find the bound $|\lambda| \lesssim 10^{88}$. This bound is not better than the bound $|\lambda| \ll 10^{90}$ that was previously obtained from the earth-based experiments \cite{Barrow:2012ay}. We then studied neutron stars and white dwarfs and we found $|\lambda| \lesssim 10^{75}$ and $|\lambda| \lesssim 10^{84}$ respectively, which are much better than the previous bound. Since the neutron mass and the proton mass depend on the QCD effective coupling constant, these bounds apply to the Horndeski term for the QCD sector of the SM. In this regard, we found the strongest bound from neutron stars. Secondly, we considered modification of the gravitational Poisson equation. Then from neutron stars with strong magnetic fields, we found the bound $|\lambda| \lesssim 10^{98}$ which is weak in comparison with the bounds that we obtained from the modification of gauge forces in the vicinity of compact astronomical objects. In the last step, we investigated the propagation of gravitational waves (GWs). We found corrections induced by the Horndeski term on the speed of GWs and speed of electromagnetic waves. We then used the GWs observational bound on the speed of GWs and we found very weak bound $|\lambda| \lesssim 10^{119}$. The Horndeski term also induces an effective mass term for the gravitons which is much suppressed in comparison with the mass term provided by the free Yang-Mills term for the system under consideration. Therefore, as far as we impose the bounds from compact astronomical objects, the Horndeski non-minimal coupling is consistent with the recent LIGO observations of GWs.

\vspace{.05cm}
{\bf Acknowledgments:} A. A. would like to thank  the Yukawa Institute for Theoretical Physics in Kyoto for hospitality and support. M. A. G thanks F. Hajkarim for useful discussions. The work of M. A. G. was supported by Japan Society for the Promotion of Science (JSPS) Grants-in-Aid for international research fellow No. 19F19313. The work of S. M. was supported by Japan Society for the Promotion of Science (JSPS) Grants-in-Aid for Scientific Research (KAKENHI) No. 17H02890, No. 17H06359, and by World Premier International Research Center Initiative (WPI), MEXT, Japan. 
\vspace{0.5cm}

{}


\begin{thebibliography}{}

\bibitem{Aad:2012tfa} 
G.~Aad {\it et al.} [ATLAS Collaboration],
Phys.\ Lett.\ B {\bf 716}, 1 (2012)
[arXiv:1207.7214 [hep-ex]].

\bibitem{Chatrchyan:2012xdj} 
S.~Chatrchyan {\it et al.} [CMS Collaboration],
Phys.\ Lett.\ B {\bf 716}, 30 (2012)
[arXiv:1207.7235 [hep-ex]].

\bibitem{Bezrukov:2007ep} 
F.~L.~Bezrukov and M.~Shaposhnikov,
Phys.\ Lett.\ B {\bf 659}, 703 (2008)
[arXiv:0710.3755 [hep-th]].

\bibitem{Barvinsky:2008ia} 
A.~O.~Barvinsky, A.~Y.~Kamenshchik and A.~A.~Starobinsky,
JCAP {\bf 0811}, 021 (2008)
[arXiv:0809.2104 [hep-ph]].

\bibitem{Bezrukov:2008ej} 
F.~L.~Bezrukov, A.~Magnin and M.~Shaposhnikov,
Phys.\ Lett.\ B {\bf 675}, 88 (2009)
[arXiv:0812.4950 [hep-ph]].

\bibitem{Rubio:2018ogq} 
J.~Rubio,
Front.\ Astron.\ Space Sci.\  {\bf 5}, 50 (2019)
[arXiv:1807.02376 [hep-ph]].

\bibitem{Horndeski:1976gi} 
G.~W.~Horndeski,
J.\ Math.\ Phys.\  {\bf 17}, 1980 (1976).

\bibitem{Davydov:2015epx} 
E.~Davydov and D.~Gal'tsov,
Phys.\ Lett.\ B {\bf 753}, 622 (2016)
[arXiv:1512.02164 [hep-th]].

\bibitem{Turner:1987bw} 
M.~S.~Turner and L.~M.~Widrow,
Phys.\ Rev.\ D {\bf 37}, 2743 (1988).

\bibitem{Mukohyama:2016npi} 
  S.~Mukohyama,
  Phys.\ Rev.\ D {\bf 94}, no. 12, 121302 (2016)
  [arXiv:1607.07041 [hep-th]].

\bibitem{Heisenberg:2018vsk} 
L.~Heisenberg,
Phys.\ Rept.\  {\bf 796}, 1 (2019)
[arXiv:1807.01725 [gr-qc]].

\bibitem{deFelice:2017paw} 
A.~de Felice, L.~Heisenberg and S.~Tsujikawa,
Phys.\ Rev.\ D {\bf 95}, no. 12, 123540 (2017)
[arXiv:1703.09573 [astro-ph.CO]].

\bibitem{DeFelice:2016uil} 
A.~De Felice, L.~Heisenberg, R.~Kase, S.~Mukohyama, S.~Tsujikawa and Y.~l.~Zhang,
Phys.\ Rev.\ D {\bf 94}, no. 4, 044024 (2016)
[arXiv:1605.05066 [gr-qc]].

\bibitem{Heisenberg:2017hwb} 
L.~Heisenberg, R.~Kase, M.~Minamitsuji and S.~Tsujikawa,
JCAP {\bf 1708}, 024 (2017)
[arXiv:1706.05115 [gr-qc]].

\bibitem{Feng:2016vyi} 
K.~Feng,
arXiv:1611.09032 [gr-qc].

\bibitem{Jimenez:2013qsa} 
J.~Beltran Jimenez, R.~Durrer, L.~Heisenberg and M.~Thorsrud,
JCAP {\bf 1310}, 064 (2013)
[arXiv:1308.1867 [hep-th]].

\bibitem{BeltranJimenez:2017cbn} 
J.~Beltran Jimenez, L.~Heisenberg, R.~Kase, R.~Namba and S.~Tsujikawa,
Phys.\ Rev.\ D {\bf 95}, no. 6, 063533 (2017)
[arXiv:1702.01193 [hep-th]].

\bibitem{Barrow:2012ay} 
J.~D.~Barrow, M.~Thorsrud and K.~Yamamoto,
JHEP {\bf 1302}, 146 (2013)
[arXiv:1211.5403 [gr-qc]].

\bibitem{HaTh}
J. B. Hartle and K. S. Thorne, 
Astrophys. J. {\bf 153}, 807-834 (1968).

\bibitem{LyndenBell:1969yx} 
D.~Lynden-Bell,
Nature {\bf 223}, 690 (1969).

\bibitem{Kormendy:1995er} 
J.~Kormendy and D.~Richstone,
Ann.\ Rev.\ Astron.\ Astrophys.\  {\bf 33}, 581 (1995).


\bibitem{Synge:1966okc} 
J.~L.~Synge,
Mon.\ Not.\ Roy.\ Astron.\ Soc.\  {\bf 131}, no. 3, 463 (1966).

\bibitem{Bardeen}
 J. M. Bardeen, inBlack Holes (Les Astres Occlus), edited by C. Dewitt and B. S. Dewitt (Gordon and Breach,New York, 1973), pp. 215–239.

\bibitem{Dokuchaev:2019jqq} 
V.~I.~Dokuchaev and N.~O.~Nazarova,
arXiv:1911.07695 [gr-qc].

\bibitem{Akiyama:2019cqa} 
K.~Akiyama {\it et al.} [Event Horizon Telescope Collaboration],
Astrophys.\ J.\  {\bf 875}, no. 1, L1 (2019)
[arXiv:1906.11238 [astro-ph.GA]].

\bibitem{Akiyama:2019eap} 
K.~Akiyama {\it et al.} [Event Horizon Telescope Collaboration],
Astrophys.\ J.\  {\bf 875}, no. 1, L6 (2019)
[arXiv:1906.11243 [astro-ph.GA]].

\bibitem{Akiyama:2019fyp} 
K.~Akiyama {\it et al.} [Event Horizon Telescope Collaboration],
Astrophys.\ J.\  {\bf 875}, no. 1, L5 (2019)
[arXiv:1906.11242 [astro-ph.GA]].

\bibitem{Akiyama:2019bqs} 
K.~Akiyama {\it et al.} [Event Horizon Telescope Collaboration],
Astrophys.\ J.\  {\bf 875}, no. 1, L4 (2019)
[arXiv:1906.11241 [astro-ph.GA]].

\bibitem{Akiyama:2019sww} 
K.~Akiyama {\it et al.} [Event Horizon Telescope Collaboration],
Astrophys.\ J.\  {\bf 875}, no. 1, L3 (2019)
[arXiv:1906.11240 [astro-ph.GA]].

\bibitem{Akiyama:2019brx} 
K.~Akiyama {\it et al.} [Event Horizon Telescope Collaboration],
Astrophys.\ J.\  {\bf 875}, no. 1, L2 (2019)
[arXiv:1906.11239 [astro-ph.IM]].

\bibitem{Allahyari:2019jqz} 
A.~Allahyari, M.~Khodadi, S.~Vagnozzi and D.~F.~Mota,
JCAP {\bf 2002}, 003 (2020)
[arXiv:1912.08231 [gr-qc]].

\bibitem{Bambi:2019tjh} 
C.~Bambi, K.~Freese, S.~Vagnozzi and L.~Visinelli,
Phys.\ Rev.\ D {\bf 100}, no. 4, 044057 (2019)
[arXiv:1904.12983 [gr-qc]].

\bibitem{Shapiro:1983du} 
S.~L.~Shapiro and S.~A.~Teukolsky,
New York, USA: Wiley (1983) 645 p

\bibitem{Ishak:2018his}
M.~Ishak,
Living Rev.\ Rel.\  {\bf 22}, no. 1, 1 (2019)
[arXiv:1806.10122 [astro-ph.CO]].

\bibitem{Landau:2020vkr} 
S.~J.~Landau,
arXiv:2002.00095 [astro-ph.CO].

\bibitem{Magano:2017mqk} 
D.~M.~N.~Magano, J.~M.~A.~Vilas Boas and C.~J.~A.~P.~Martins,
Phys.\ Rev.\ D {\bf 96}, no. 8, 083012 (2017)
[arXiv:1710.05828 [astro-ph.CO]].

\bibitem{Jain:2015edg} 
R.~K.~Jain, C.~Kouvaris and N.~G.~Nielsen,
Phys.\ Rev.\ Lett.\  {\bf 116}, no. 15, 151103 (2016)
[arXiv:1512.05946 [astro-ph.CO]].

\bibitem{Uzan:2010pm} 
J.~P.~Uzan,
Living Rev.\ Rel.\  {\bf 14}, 2 (2011)
[arXiv:1009.5514 [astro-ph.CO]].

\end{thebibliography}
\end{document}